\documentclass[twocolumn,twocolappendix]{aastex631}
\usepackage{amsmath,amssymb,graphicx}
\usepackage{epsf,verbatim}
\usepackage{hyperref}
\usepackage{comment}
\usepackage{color}
\usepackage{slashed}
\usepackage{subfigure}
\usepackage{comment}
\usepackage{mdwlist, paralist}
\usepackage{rotating}
\usepackage{bm}
\usepackage{multirow}

\newcommand{\beq}{\begin{equation}}
\newcommand{\eeq}{\end{equation}}
\newcommand{\bal}{\begin{align}}
\newcommand{\eal}{\end{align}}
\newcommand{\bit}{\begin{itemize}}
\newcommand{\eit}{\end{itemize}}
\newcommand{\ben}{\begin{enumerate}}
\newcommand{\een}{\end{enumerate}}

\renewcommand{\eqref}[1]{Eq.~(\ref{#1})}

\newcommand{\figref}[1]{Fig.~\ref{fig:#1}}

\renewcommand{\t}{\tilde}

\newcommand{\f}{\frac}

\newcommand{\msun}{\,\rm M_\odot}

\begin{document}

\title{Seeding Supermassive Black Holes with Self-Interacting Dark Matter: \\
A Unified Scenario with Baryons}

\author[0000-0002-9048-2992]{Wei-Xiang Feng}
\email{wfeng016@ucr.edu}
\affiliation{Department of Physics and Astronomy, University of California, Riverside, CA 92521, USA}

\author[0000-0002-8421-8597]{Hai-Bo Yu}
\email{haiboyu@ucr.edu}
\affiliation{Department of Physics and Astronomy, University of California, Riverside, CA 92521, USA}

\author[0000-0001-9922-6162]{Yi-Ming Zhong}

\email{ymzhong@kicp.uchicago.edu}
\affiliation{Kavli Institute for Cosmological Physics, University of Chicago, Chicago, IL 60637, USA}

\begin{abstract}
Observations show that supermassive black holes (SMBHs) with a mass of $\sim10^9\msun$ exist when the Universe is just $6\%$ of its current age. We propose a scenario where a self-interacting dark matter halo experiences gravothermal instability and its central region collapses into a seed black hole. The presence of baryons in protogalaxies could significantly accelerate the gravothermal evolution of the halo and shorten collapse timescales. The central halo could dissipate its angular momentum remnant via viscosity induced by the self-interactions. The host halo must be on high tails of density fluctuations, implying that high-$z$ SMBHs are expected to be rare in this scenario. We further derive conditions for triggering general relativistic instability of the collapsed region. Our results indicate that self-interacting dark matter can provide a unified explanation for diverse dark matter distributions in galaxies today and the origin of SMBHs at redshifts $z\sim6\textup{--}7$.

\end{abstract}


\section{Introduction}
Astrophysical observations of high-redshift quasars indicate that $\sim10^9\msun$ black holes exist when the Universe is just $800~{\rm Myr}$ old after the Big Bang ($z\sim7$), see~\cite{Inayoshi:2019fun} for a review. The origin of these supermassive black holes (SMBHs) is still a mystery. In particular, it is extremely puzzling how they could become so massive in such a short time. A popular idea is that there exist heavy seed black holes in the early Universe and they grow massive by accreting baryons~\citep{Volonteri:2010wz,Natarajan2011}. Assuming Eddington accretion, we can relate the black hole mass ($M_\text{BH}$) and its seed mass ($M_{\rm seed}$) as~\citep{Salpeter:1964kb}
\begin{equation}
M_\text{BH}=M_{\rm seed}\exp(\Delta t/\tau),
\label{eq:growth}
\end{equation}
where $\Delta t$ is the elapse time and  $\tau =(450/f_\text{Edd}) [\epsilon_M/(1-\epsilon_M)]~{\rm Myr}$ is the $e$-folding time. $\epsilon_M$ is the radiative efficiency and commonly assumed to be $0.1$~\citep{Shakura:1976xk}, and $f_{\rm Edd}$ is the Eddington ratio following $f_{\rm Edd} = L_{\rm bol}/L_{\rm Edd}$, where $L_{\rm bol}$ is the observed bolometric luminosity and $L_{\rm Edd} = 1.3 \times10^{38}(M_{\rm BH}/{\rm M_\odot})~{\rm erg/s}$ is the Eddington luminosity. $\epsilon_M$ measures the efficiency of conversion of mass energy to luminous energy by accretion, while $f_{\rm Edd}$ characterizes the efficiency of accretion luminosity.    

Consider J1007+2115, the most massive known quasar with $M_\text{BH}\approx1.5\times10^9 \msun$ at $z> 7.5$~\citep{yang2020poniua}.  Taking $f_\text{Edd}\simeq 1$, we estimate $M_{\rm seed}\sim 10^4\msun$ if it forms at $z\sim30$, i.e., $\Delta t=597~{\rm Myr}$ to its observed $z=7.51$. Such a seed is too massive to be produced from collapsed Pop III stars~\citep{Inayoshi:2019fun}, but it could form through the direct collapse of pristine baryonic gas~\citep{Bromm:2002hb,Begelman:2006db}; see also~\cite{Freese:2010re}. The latter scenario predicts $M_{\rm seed}\sim10^{5}\textup{--}10^6\msun$. However, observations show there are high-$z$ SMBHs with $f_\text{Edd}$ much less than $1$~\citep{onoue2019subaru,matsuoka2019discovery}. For example, J1205-0000 is observed at $z=6.7$ with $M_\text{BH}=2.2\times 10^9\msun$ and $f_\text{Edd} = 0.16$~\citep{onoue2019subaru}. The Eddington accretion then implies it grows from a seed with a mass of $2\times10^8 \msun$ at $z\sim 30$, too heavy to be produced via the direct collapse of gas.

There could be complications in those estimates. For example, the accretion luminosity may not be a constant over time, and SMBHs could experience a rapid accretion phase beyond the Eddington limit~\citep{Begelman:1979,Volonteri:2005fj,Alexander:2014noa}. In addition, the radiative efficiency depends on black hole spin. For standard thin disks, which drive the hole to maximal spin, $\epsilon_M\sim0.42$, while for magnetohydrodynamic disks $\epsilon_M\sim0.2$~\citep{Shapiro:2004ud} or even lower $\epsilon_M\sim0.01$ associated with super-Eddington accretion~\citep{McKinney:2013txa,Sadowski:2015}. Furthermore, mergers could amplify the black hole mass. It's possible that the existence of SMBHs can be explained after taking into account these effects, but more work is needed to understand how they affect individual ones; see~\cite{Inayoshi:2019fun} for more discussion.

In this {\em Letter}, we study the scenario of gravothermal collapse of self-interacting dark matter (SIDM)~\citep{Spergel:1999mh,Kaplinghat:2015aga,Tulin:2017ara} in explaining the origin of high-$z$ SMBHs. Dark matter self-interactions can transport heat in the halo over cosmological timescales~\citep{Dave:2000ar,Ahn:2004xt,Rocha:2012jg,Vogelsberger:2012ku}. As a gravothermal system, the SIDM halo has negative heat capacity~\citep{LyndenBell:1968yw}. The central region could become hot and collapse to a singular state with a finite mass at late stages of the evolution~\citep{Balberg:2002ue,Balberg:2001qg}. Thus SIDM has a natural mechanism in triggering dynamical instability, a necessary condition to form a black hole. Recent studies also show that SIDM is favored for explaining diverse dark matter distributions over a wide range of galactic systems, see~\cite{Tulin:2017ara} for a review. It is intriguing to explore an SIDM scenario that may explain the origin of the high-$z$ SMBHs and observations of galaxies at $z\sim0$.

We adopt a typical baryon mass profile for high-$z$ protogalaxies, and show the collapse time can be shortened by a factor of $100$, compared to the SIDM-only case. Even for the self-scattering cross section per unit mass $\sigma/m\sim1~{\rm cm^2/g}$, broadly consistent with the value used to explain galactic observations~\citep{Tulin:2017ara}, the central halo could collapse sufficiently fast to form a  seed for $z\gtrsim7$. Depending on the halo mass, this scenario could explain the origin of high-$z$ SMBHs with $f_\text{Edd}\sim1$ and $0.1$. It also has a built-in mechanism to dissipate angular momentum remanent of the central halo, i.e., viscosity induced by the self-interactions. We will further show when the 3D velocity dispersion of SIDM particles in the collapsed central region reaches $0.57c$, the general relativistic (GR) instability can be triggered. We demonstrate a unified SIDM scenario that could explain observations of galaxies today and high-$z$ SMBHs. In the appendices, we provide additional information.

\section{Gravothermal evolution} 

We use a conducting fluid model~\citep{Balberg:2002ue,Koda:2011yb} to study the gravothermal evolution of an SIDM halo, as it yields high resolution for us to closely trace the collapse process. To capture the influence of baryons, we extend the original model with a baryonic component. The evolution of the halo can be described by the following equations
\begin{eqnarray}
\frac{\partial M_{\chi}}{\partial r} = 4\pi r^2 \rho_\chi,\;
\frac{\partial (\rho_\chi \nu_\chi^2)}{\partial r} = - \frac{G (M_\chi+M_{b}) \rho_\chi}{r^2},\nonumber\\
\frac{\partial L_\chi}{\partial r} = -4 \pi \rho_\chi r^2 \nu_\chi^2 D_t \ln \frac{\nu_\chi^3}{\rho_\chi},\;
\frac{L_\chi}{4\pi r^2}=-\kappa\frac{\partial (m\nu^2_\chi)}{k_B\partial r},
\end{eqnarray}
where $M_\chi, \rho_\chi, \nu_\chi$, and $L_\chi$ are dark matter mass, density, 1D velocity dispersion, and luminosity profiles, respectively, and they are dynamical variables and evolve with time; $M_{b}$ is the baryon mass profile in the host galaxy; $k_B$ is the Boltzmann constant, $G$ is the Newton constant, and $D_t$ denotes the Lagrangian time derivative. Heat conductivity of the dark matter fluid $\kappa$ can be expressed as $\kappa=(\kappa^{-1}_{\rm lmfp}+\kappa^{-1}_{\rm smfp})^{-1}$, where $\kappa_{\rm lmfp}\approx0.27C \rho_\chi\nu^3_\chi\sigma k_B/(Gm^2)$ and $\kappa_{\rm smfp}\approx2.1\nu_\chi k_B/\sigma$ denote conductivity in the long- and short-mean-free-path regimes, respectively, and we set $C\simeq0.75$ based on calibrations with N-body simulations~\citep{Pollack:2014rja, Essig:2018pzq}. In the short-mean-free-path regime, heat conduction can be characterized by the self-interaction mean free path $\lambda = m/\rho_\chi\sigma$ and $Kn=\lambda/H <1$, where $H =(\nu_\chi^2/4\pi G \rho_\chi)^{1/2}$ is the scale height. In the long-mean-free-path regime, it's characterized by $H$ and $Kn >1$.

We assume the {\it initial} halo follows a Navarro-Frenk-White (NFW) profile~\citep{Navarro:1995iw} with $r_s$ and $\rho_s$ as its scale radius and density, respectively. The boundary conditions are $M_\chi=0$ at $r=0$, $M_\chi=M_{200}$ and $L_\chi=0$ at $r=r_{200}$, where $M_{200}$ and $r_{200}$ are the virial halo mass and radius, respectively; they are are equivalent to $r_s$ and $\rho_s$ in specifying a halo for a given redshift $z$. We adopt the baryon mass profile $M_b (r)\approx 0.1 (4\pi \rho_s r_s^3) (r/r_s)^{0.6}$, based on cosmological hydrodynamical simulations of protogalaxies at $z\sim17$~\citep{Wise:2007bf}; see Appendix A. As an approximation, we assume the baryon mass profile is static and it does not evolve with time. We recast the fluid equations with dimensionless variables and solve them numerically using the method as in~\cite{Balberg:2002ue,Pollack:2014rja,Essig:2018pzq}. The fiducial quantities relevant for later discussions are $M_0=4\pi\rho_s r^3_s$, $t_0=1/\sqrt{4\pi G\rho_s}$ and $(\sigma/m)_0=1/(r_s\rho_s)$; hence $M_b (r)= 0.1 M_0 (r/r_s)^{0.6}$. We then map dimensionless outputs from the simulations to physical ones assuming Planck cosmology, i.e., $h = 0.67$, $\Omega_m = 0.315$, and $\Omega_\Lambda = 0.685$~\citep{Aghanim:2018eyx}.

We further elaborate our assumptions made above. The initial halo is optically thin at its characteristic radius  if $(\sigma/m)(r_s\rho_s)<1$~\citep{Pollack:2014rja}. Our numerical study takes $(\sigma/m)(r_s\rho_s)=0.5$, and hence an NFW initial condition is well justified. As a concrete example, we take the simulated baryon profile from~\cite{Wise:2007bf}, which is based on collisionless dark matter. In SIDM, the baryon profile could be diffuse because of halo core formation~\citep{Vogelsberger:2014vr,Cruz:2020rit}. However, if baryon infall occurs early before a large core forms, the baryon distribution can be as compact as the one predicted in the collisionless limit~\citep{Robertson:2017mgj}, or even more dense if SIDM collapse occurs~\citep{Sameie:2021ang}, albeit both simulations focus on systems at low redshifts. 

Our approximation of a static baryon mass profile could be conservative in estimating the collapse time as the baryons would further contract when the collapse starts, see~\cite{Sameie:2021ang}. We have also used a Hernquist profile~\citep{Hernquist:1990be} to model the baryon distribution and obtained similar results if the baryons dominate the central potential as the simulated galaxies in~\cite{Wise:2007bf}. Given these considerations, our model assumptions are justified. Nevertheless, it would be of interest to simulate high-$z$ protogalaxies in SIDM and test our assumptions in a cosmological environment.

\begin{figure}[htbp]
   \centering
   \includegraphics[width=0.4\textwidth]{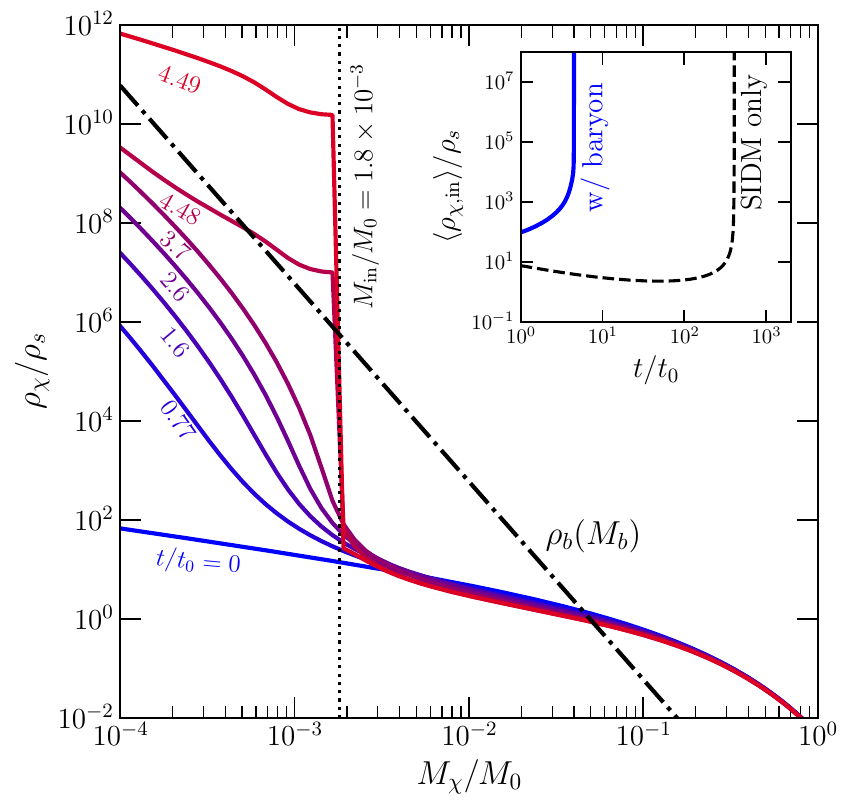} 
   \caption{Gravothermal evolution of the dark matter density vs. enclosed mass in the presence of the baryonic potential (solid), as well as the fixed baryon profile (dash-dotted). Each dark matter profile is labeled with its corresponding evolution time, and the vertical dotted line indicates the mass of the central halo that would eventually collapse into a seed black hole. The {\em insert} panel illustrates the evolution of the averaged dark matter density of the central halo with (solid) and without (dashed) including the baryons.}
   \label{fig:evolution}
\end{figure}

\begin{figure}[htbp]
   \centering
\includegraphics[width=0.45\textwidth]{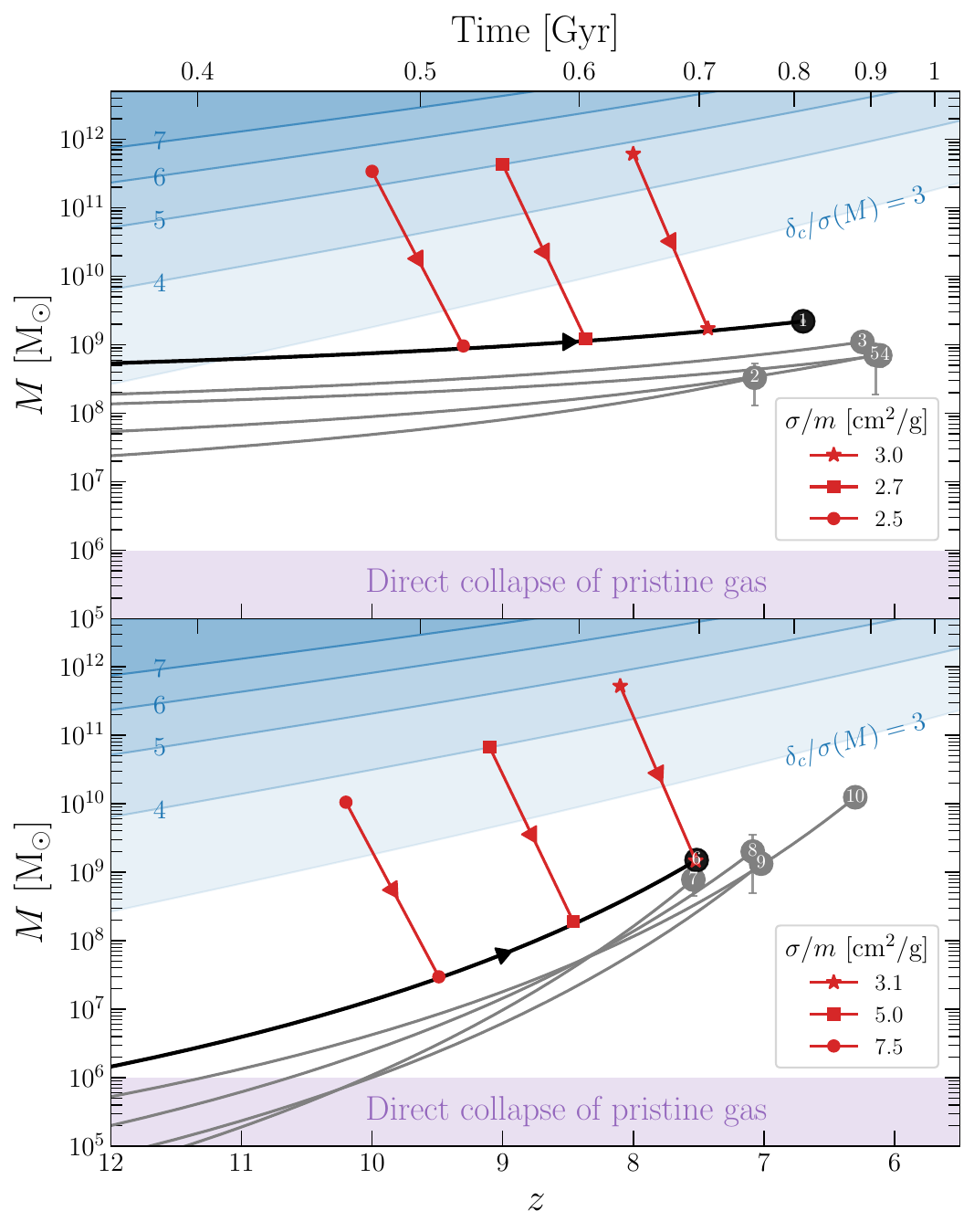}
   \caption{SIDM benchmarks (red) that could explain the origin of the SMBHs J1205-0000 (labeled as ``1", upper panel) and J1007+2115 (``6", lower panel) with an observed Eddington growth rate of $f_\text{Edd} =0.16$~\citep{onoue2019subaru} and $1.06$~\citep{yang2020poniua}, respectively. The black curves indicate their Eddington accretion history. For each red arrow, the markers on higher and lower $z$ end denote initial halo and seed masses, respectively, and the horizontal difference between the two ends indicates the timescale of gravothermal collapse. The blue shaded regions indicate the ratio of the critical density fluctuation to the halo mass variance. The magenta bands denote the mass range of the seed produced via the direct collapse of pristine gas. The gray curves are Eddington growth history of other high-$z$ SMBHs with $f_\text{Edd}\sim 0.1$ (upper) and $\sim1$ (lower).}
   \label{fig:cosmo}
\end{figure}

\section{Roles of baryons} 

Fig.~\ref{fig:evolution} shows the gravothermal evolution of the dark matter density vs. enclosed mass (solid) in the presence of the baryons (dash-dotted), where we fix $(\sigma/m)(r_s\rho_s)=0.5$. The insert panel illustrates the average inner density vs. evolution time with (solid) and without (dashed) including the baryon mass. The average inner density $\left<\rho_{\chi, {\rm in}}\right>$ is calculated within the central region where the enclosed mass equals to that of the seed black hole, as we will explain later. With the baryons, the halo does not form a large density core and it quickly evolves into the collapse phase~\citep{Sameie:2018chj,Yang:2021kdf}. Its density keeps increasing and eventually becomes super-exponential in the end. The collapse timescale is $t_c=4.49 t_0$, a factor of $\sim 90$ shorter than the one predicted in the SIDM-only case with the same interaction strength. We also performed simulations with a Hernquist baryon profile and found a similar reduction factor in $t_c$ if the baryon distribution is as compact as that in~\cite{Wise:2007bf}. 

We also see that as the central density increases for $t\gtrsim4.48 t_0$, the enclosed mass for a central region remains almost a constant $M_\text{in} \approx 1.8 \times10^{-3} M_0$. This is the region where the halo is in the short-mean-free-path regime. A similar phenomenon also occurs without including the baryons~\citep{Balberg:2002ue}. For the SIDM-only case we consider, the corresponding $M_\text{in}/M_0$ value is $4.2 \times10^{-2}$, which is larger than the one with the baryons. As the halo evolves further, the density continues increasing and the central halo ($Kn\lesssim1$) would eventually collapse into a singular state, a seed black hole. We assume the seed mass $M_\text{seed}=M_\text{in}$, suggested by numerical studies of collapsed massive stars~\citep{Saijo:2002qt}.

\section{Seeding Supermassive Black Holes}

To explain the origin of high-$z$ SMBHs, the initial halo must be sufficiently heavy {\em and} collapse fast enough. We first check the scaling relations $M_\text{in}\propto M_0\propto M_{200}$, and $t_{\rm c}\propto r_{s}^{-1}\rho_{s}^{-3/2}\propto M_{200}^{-1/3}c_{200}^{-7/2}(1+z)^{-7/2}$~\citep{Essig:2018pzq}, where $c_{200}=r_{200}/r_s$ is the halo concentration. Apparently, $t_c$ is very sensitive to $c_{200}$. There is a tight correlation between $c_{200}$ and $M_{200}$ for halos at $z\lesssim5$, but the $c_{200}$ distribution at higher redshifts is less known. There is a trend that $c_{200}$ gradually becomes independent of $M_{200}$ and its median asymptote to $c_{200}\sim3$ at $z\sim5\textup{--}10$~\citep{Dutton:2014xda,Zhao:2008wd}. We fix $c_{200}=3$, and leave with two parameters $M_{200}$ and $z$ to vary. 

Fig.~\ref{fig:cosmo} shows benchmarks (red) that could explain the origin of the SMBHs J1205-0000 with the Eddington ratio $f_\text{Edd} =0.16$~\citep{onoue2019subaru} (upper panel) and J1007+2115 with $f_\text{Edd}=1.06$~\citep{yang2020poniua} (lower panel). The black curves indicate their Eddington accretion history reconstructed using Eq.~\ref{eq:growth}. For reference, the gray ones denote the accretion history of other high-$z$ SMBHs with $f_\text{Edd}\sim0.1$ (upper)~\citep{onoue2019subaru,matsuoka2019discovery} and those $f_\text{Edd}\sim1$ (lower)~\citep{yang2020poniua,Banados:2017unc,mortlock2011luminous,wang2018discovery,wu2015ultraluminous}. We have checked all of them (gray) could also be explained within our scenario. The direct collapse of pristine gas could provide massive enough seeds (magenta) for the SMBHs with $f_\text{Edd}\sim1$, but not those with $f_\text{Edd}\sim 0.1$~\citep{onoue2019subaru}.

As the redshift of the initial halo increases, the favored halo mass becomes smaller, because the seed black hole has more time to grow. To explain the origin of the SMBHs with $f_\text{Edd}\sim1$, the mass is in a range of $M_{200}\sim10^{9}\textup{--}10^{11}\msun$ for $z\sim11\textup{--}9$. For those with $f_\text{Edd}\sim0.1$, $M_{200}$ needs to be relatively higher, $\sim10^{11}\textup{--}10^{12}\msun$, as their growth rate is much smaller and a heavier seed is required. We have checked that the overall trend holds for halos with $z\gtrsim11$.

As an example, we take the case with $(M_{200}/{\rm M_\odot}, z)=(6.8\times 10^{11}, 8)$ that seeds J1205-0000, the most challenging SMBH, to demonstrate our derivation. For the halo, $\rho_{s}\approx8.1\times10^7\,{\rm M_\odot/kpc^3}$ and $r_{s}\approx10~{\rm kpc}$. Hence the fiducial parameters are $t_0=15~{\rm Myr}$, $M_0=1.1\times10^{12}\msun$ and $(\sigma/m)_0=5.8~{\rm cm^2/g}$. The seed mass is $M_{\rm in}=1.8\times10^{-3}M_0\approx1.9\times10^9\msun$ and the collapse time $t_c{=4.5t_0}\approx68~{\rm Myr}$, and the self-scattering cross section $\sigma/m\approx3.0~{\rm cm^2/g}$. Since $z=8$ is equivalent to $t=642~{\rm Myr}$ after the Big Bang, the seed forms at $710~{\rm Myr}$ ($z=7.4$). For an SIDM-only halo with the same parameters, we find $t_{c}\approx400 t_0\approx6~{\rm Gyr}$, too long to form a seed. 

To speed up the collapse process in the absence of the baryonic influence, one would need to take much larger $\sigma/m$ and $c_{200}$~\citep{Pollack:2014rja}, or consider dissipative self-interactions~\citep{Choquette:2018lvq,Essig:2018pzq,Huo:2019yhk}. For comparison, our scenario predicts $\sigma/m\sim1\textup{--}10~{\rm cm^2/g}$ within a minimal elastic SIDM model that has been shown to explain dark matter distributions in the spirals~\citep{Ren:2018jpt}, Milky Way satellites~\citep{Sameie:2019zfo}, and dark-matter-deficient galaxies~\citep{Yang:2020iya}. It's important to note dwarf galaxies at present that favor a large density core are those with diffuse baryon distributions~\citep{Kaplinghat:2019dhn}. Thus their host SIDM halos would still be in the core-expansion phase and a shallow density profile is expected. In addition, in many well-motivated particle physics realizations of SIDM, see~\cite{Tulin:2017ara}, the cross section diminishes towards cluster scales. Thus, the stringent bounds on $\sigma/m$ from galaxy clusters~\citep{Kaplinghat:2015aga,Andrade:2020lqq} can be avoided. 

\section{Density fluctuations}

For the benchmark cases, the halo mass is in the range of $10^{9}\textup{--}10^{12}\msun$ for $z\sim11\textup{--}8$. We use the standard Press-Schechter formalism~\citep{Press:1973iz} to examine conditions for realizing those halos in the early Universe. The halo mass function scales as ${dn(M,z)}/{dM} \propto\exp[-\delta^2_{\rm c}(z)/2\sigma^2(M)]$~\citep{mo2010galaxy}, where $\delta_{c}(z)$ is the critical density fluctuation at $z$ and $\sigma(M)$ the mass variance. We shaded the regions with various values of $\delta_c (z)/\sigma (M)$ in Fig.~\ref{fig:cosmo} (blue). As expected, the halo mass increases as the density fluctuation increases and more massive halos form at later times. 

The halos for seeding the SMBHs with a sub-Eddington (Eddington) accretion rate correspond to $\delta_c (z)/\sigma (M)\sim4\textup{--}6 $ ($3\textup{--}5$). In addition, the baryon concentration of host galaxies needs to be high as well such that the gravothermal collapse could occur fast enough. Thus our scenario predicts that high-$z$ SMBHs should be rare. Indeed, observations show they are extremely rare in the Universe. Quasar surveys indicate that the number density of luminous ($L_\text{AGN}\gtrsim10^{46}\,\text{erg}/\text{s}$) high-$z$ SMBHs with $M_{\rm BH}\sim10^9\msun$ is $\lesssim10^{-7}~\text{Mpc}^{-3}$~\citep{Kulkarni:2018ebj,Inayoshi:2019fun,Trakhtenbrot:2020ugp}. The ratio of their mass to the dynamical (gas+stars) mass is $M_{\rm BH}/M_{\rm b}\sim1/100\textup{--}1/30$~\citep{Trakhtenbrot:2020ugp}. Taking $M_{\rm b}/M_{200}\sim0.2$, we find $M_{\rm BH}/M_{200}\sim(2\textup{--}7)\times10^{-3}$, broadly consistent with our prediction. We also note that baryon infall can occur for a halo heavier than $5\times10^3 [(1+z)/10]^{1.5}\msun$~\citep{mo2010galaxy}, and all of the benchmarks satisfy this condition easily.

\begin{figure}[t!]
   \centering
   \includegraphics[width=0.45\textwidth]{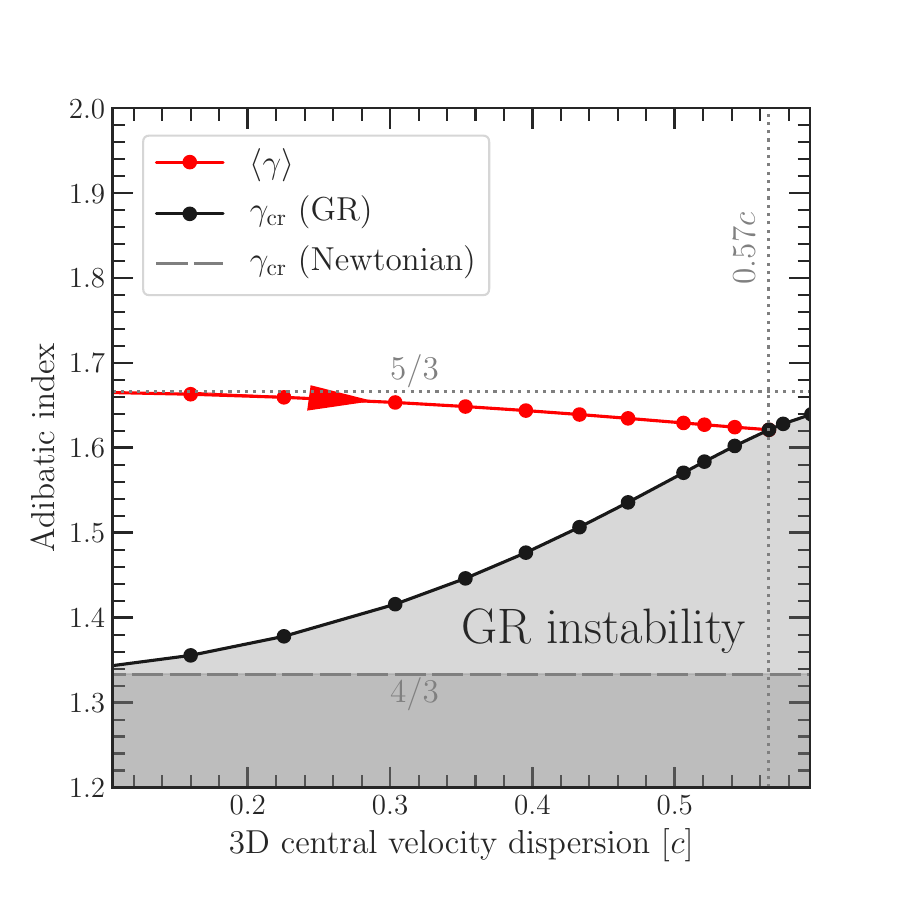} 
   \caption{The pressure-averaged adiabatic index $\left<\gamma\right>$ (red) and the critical index $\gamma_\text{cr}$ (black) vs. the central 3D velocity dispersion for each GR configuration (dot). When $\left<\gamma\right><\gamma_\text{cr}$, the system triggers the GR instability. In the Newtonian limit, $\left<\gamma\right>=5/3$ for a monatomic ideal gas, and the instability condition is $\left<\gamma\right><4/3$.
  }
   \label{fig:stability}
\end{figure}

\section{Angular momentum} 

The angular momentum remnant of the inner halo could counter gravity and even slow down the gravothermal collapse. Besides, there is an upper limit on the specific angular momentum of a black hole, $J_\text{BH}/M_\text{BH}\leq(G/c){M_\text{BH}}\approx1.4\times10^{-4}(M_\text{BH}/10^{7}\msun)~{\rm kpc\cdot km/s}$~\citep{Kerr:1963ud}. Consider the benchmark with $(M_{200}/{\rm M_\odot}, z)=(6.8\times 10^{11}, 8)$ again, dark matter particles within the radius $r_\text{in}=0.063 r_s\approx0.63~\text{kpc}$ of the initial NFW halo would collapse to a seed, where the total enclosed mass is $M_\text{in}$. We find $J_{\chi}/M_\text{in} \approx 8~{\rm kpc\cdot ~km/s}$ for the halo within $r_\text{in}$, based on a universal fitting formula~\citep{Liao:2016tdi}. This is a factor of $100$ larger than the allowed value for a $10^9\msun$ seed. 
 
Fortunately, dark matter self-interactions that lead to heat conductivity also induce viscosity, which dissipates the angular momentum remnant. In the long-mean-free-path regime, we find decays as
\begin{equation}
J_{\chi}^f\approx J_{\chi}^i \exp\left[-\frac{8}{\sqrt{27\pi}}\int dt\,\frac{\rho_\chi (\sigma/m) \nu_\chi^3 r_\text{in} }{kG {M_\chi}} \right],
\label{eq:angular}
\end{equation}
where ${J}_{\chi}^i$ and ${J}_{\chi}^{f}$ are the initial and final angular momenta of the central halo within $r_\text{in}$, respectively, and $k=2(\beta+3)/3(\beta+5)$ for a power law of $\rho_\chi\propto r^{\beta}$; see Appendix C. For an NFW (cored) profile, $\beta= -1~(0)$ and hence $k = 1/3~(2/5)$. Consider the benchmark, we have $r_\text{in}= 0.063 r_{s}$, $M_\chi=M_\text{in}=1.8\times10^{-3}M_0$, $\rho_\chi (r_\text{in})= 14 \rho_s$, $\sigma/m = 0.5/(r_s \rho_s)$ and $\nu_\chi({r_{\rm in}})\approx 0.48 \sqrt{4\pi G\rho_s}r_{s}$. Taking $k=1/3$, we find the timescale for achieving $J_{\chi}^f\sim 10^{-2} J_{\chi}^i$ is $\Delta t\approx0.1 t_0$, much shorter than that of gravothermal collapse $t_c\approx4.5 t_0$. We have checked that the other five benchmarks in Fig.~\ref{fig:cosmo} satisfy the dissipation condition. In SIDM, viscosity and conductivity share the same microscopic nature, and both effects are critical for seeding the SMBHs in our scenario. 

\section{Relativistic Instability}

As the central density increases, the velocity dispersion of the collapsed central region increases as well~\citep{Balberg:2002ue}, and it would eventually approach the relativistic limit. To see the fate of the central halo where $Kn\lesssim1$, we examine conditions for reaching GR instability. Motivated by early studies on globular cluster systems~\citep{King:1966fn,Merafina:1989}, we assume that the number density of SIDM particles in the central halo at late stages follow a truncated Maxwell-Boltzmann distribution
\begin{equation}
dn(r)\propto
\begin{cases}
(e^{-\epsilon/k_B T}-e^{-\epsilon_c/k_BT})d^3p(\epsilon)~~(\epsilon\leq\epsilon_c)\\
0\quad(\epsilon>\epsilon_c),
\end{cases}
\label{eq:mb}
\end{equation}
where $T$, $\epsilon$, and $p$ are temperature, energy, and momentum; respectively; $\epsilon_{\rm c}$ is the cutoff energy, above which the particle escapes to the boundary. Given the distribution in Eq.~\ref{eq:mb}, we use the method in~\cite{Merafina:1989} and solve the Tolman-Oppenheimer-Volkov equation to find density and pressure profiles for the collapsed central region, where we impose the boundary condition $k_BT=0.1mc^2$. For each configuration, we follow Chandrasekhar's criterion~\citep{Chandrasekhar:1964zz}, and calculate the critical adiabatic index $\gamma_{\rm cr}$ and the pressure-averaged adiabatic index $\langle\gamma\rangle$. The sufficient condition for the system to collapse into a black hole is $\langle\gamma\rangle<\gamma_{\rm cr}$. We will discuss technical details in a companion paper~\citep{fyz}.

Fig.~\ref{fig:stability} shows the averaged adiabatic index $\left<\gamma\right>$ (red) and the critical index $\gamma_\text{cr}$ (black) vs. 3D velocity dispersion at the center for each configuration denoted as a dot. As the velocity dispersion increases, its averaged index $\left<\gamma\right>$ gradually decreases from its non-relativistic limit for monatomic ideal gas $5/3$ towards the ultra-relativistic limit $4/3$. In contrast, the critical index $\gamma_\text{cr}$ increases from the Newtonian limit $4/3$~\citep{Shapiro:1983du}. This is because as the pressure starts to dominate the energy density towards the GR limit, it destabilizes the system instead. The relativistic instability occurs when the 3D central velocity dispersion approaches $0.57c$ and $\left<\gamma\right>=\gamma_{\rm cr}\approx1.62$, at which the corresponding fractional binding energy is $0.033$~\citep{fyz}.

\section{Conclusions} 

We have presented a scenario that could explain the origin of high-$z$ SMBHs with Eddington {\em and}  sub-Eddington accretion rates. The presence of baryons in protogalaxies could deepen the gravitational potential and expedite the gravothermal collapse of an SIDM halo. The favored self-scattering cross section is broadly consistent with the one used to explain diverse dark matter distributions of galaxies. In this scenario, dark matter self-interactions induce viscosity that dissipates the angular momentum remnant of the central halo. The initial halo must be on high tails of density fluctuations, which may explain why high-$z$ SMBHs are extremely rare in observations. We also checked that the GR instability condition can be satisfied. The upcoming and future facilities are expected to search for quasars with a wide range of luminosities~\citep{Trakhtenbrot:2020ugp}. The observations would provide a more complete picture of populations of high-$z$ SMBHs and further test our scenario.

\section*{Acknowledgements}
We acknowledge Alexander Dreichner for helping identify a bug in the previous version of the fluid simulation code. We thank Stuart L. Shapiro for helpful and friendly discussions, and Masafusa Onoue for clarifying the Eddington growth rate of J2216-0016. WXF acknowledges the Institute of Physics, Academia Sinica, for the hospitality during the completion of this work. This work was supported by U.S. Department of Energy under Grant No. de-sc0008541 (HBY) NASA grant 80NSSC20K0566 (HBY), and in part by the Kavli Institute for Cosmological Physics at the University of Chicago through an endowment from the Kavli Foundation and its founder Fred Kavli (YZ). This project was made possible through the support of a grant from the John Templeton Foundation (HBY, ID\# 61884). The opinions expressed in this publication are those of the authors and do not necessarily reflect the views of the John Templeton Foundation.

\appendix

\section{The gas density profile}

\begin{figure}[htbp]
   \centering
   \includegraphics[width=0.4\textwidth]{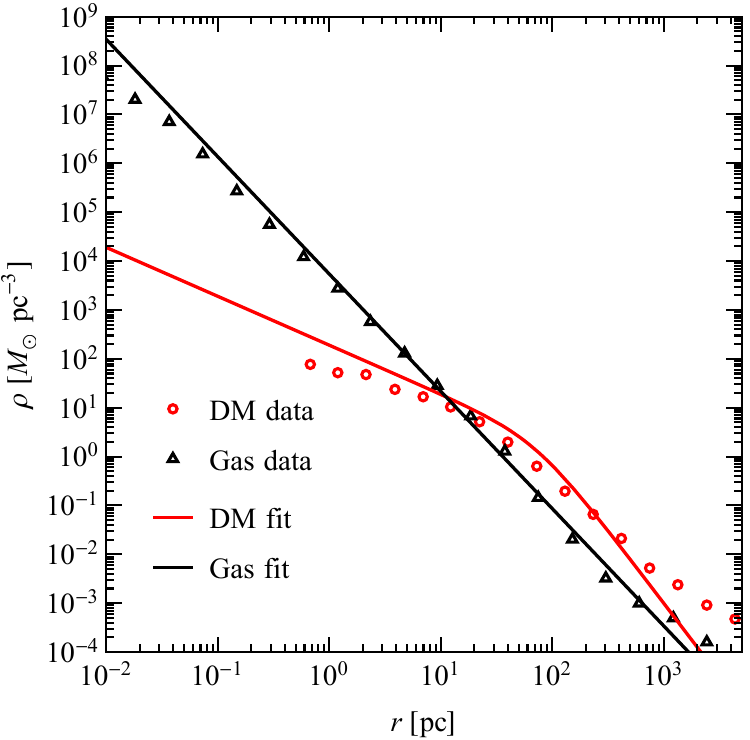} 
   \caption{Dark matter (red) and gas (black) density profiles after fitting to the simulated ones in \cite{Wise:2007bf}; see their Fig. 4 (right, panel b).}
   \label{fig:fit}
\end{figure}

To model the gas distribution of protogalaxies, we adopt simulation results in~\cite{Wise:2007bf} (simulation B). Their simulated gas and dark matter distributions are fitted with a single power law of $\rho_b\sim r^{-2.4}$ and an NFW profile, respectively. We find the following ansatz works well for the gas.
\beq
\rho_b (r) = \rho_{b,s} \left(\f{r}{r_s}\right)^{-2.4},
\eeq
where $\rho_{b,s}$ is the scale density of the gas and $r_s$ is the scale radius of the simulated halo. The corresponding mass profile is 
\beq
M_b(r) = 1.67 \times \f{\rho_{b,s}}{\rho_s} (4\pi \rho_s r_s^3) \left(\f{r}{r_s}\right)^{0.6}
\eeq
We use simulation data shown in Fig.~4 (right, panel b) in~\cite{Wise:2007bf} to fix the model parameters, $r_s = 73\,\text{pc}$, $\rho_s = 2.6\msun/\text{pc}^3$, and $\rho_{b,s}= 0.19\msun/\text{pc}^3$; see Fig.~\ref{fig:fit} for comparison. Since $1.67 \times {\rho_{b,s}}/{\rho_s}\approx 0.1$, we take $M_b(r) = 0.1 M_0 (r/r_s)^{0.6}$ for the static baryon distribution in our semi-analytical simulations, shown as the dash-dotted line in the left panel of Fig.~\ref{fig:mass}. Note that the results from~\cite{Wise:2007bf} have high enough resolutions for setting initial conditions in our simulations, where we trace the collapse process with the conducting fluid model.

\section{The numerical procedure}

\begin{figure*}[htbp]
   \centering
   \includegraphics[width=0.4\textwidth]{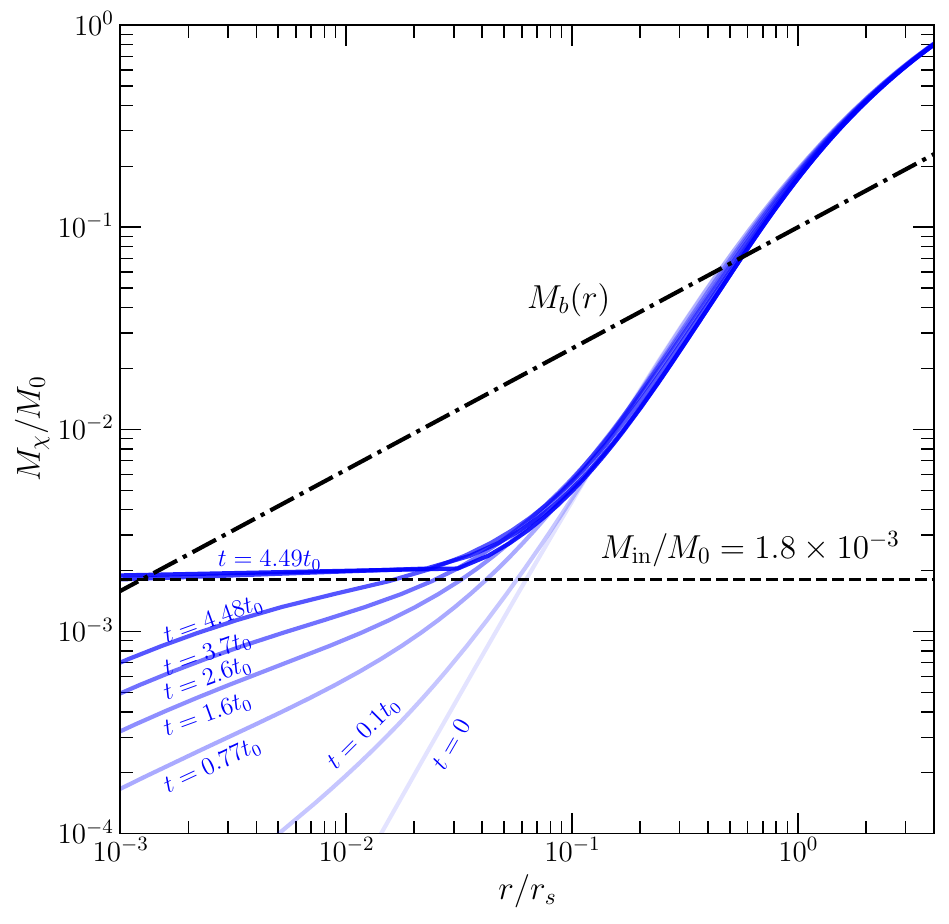}
   \includegraphics[width=0.4\textwidth]{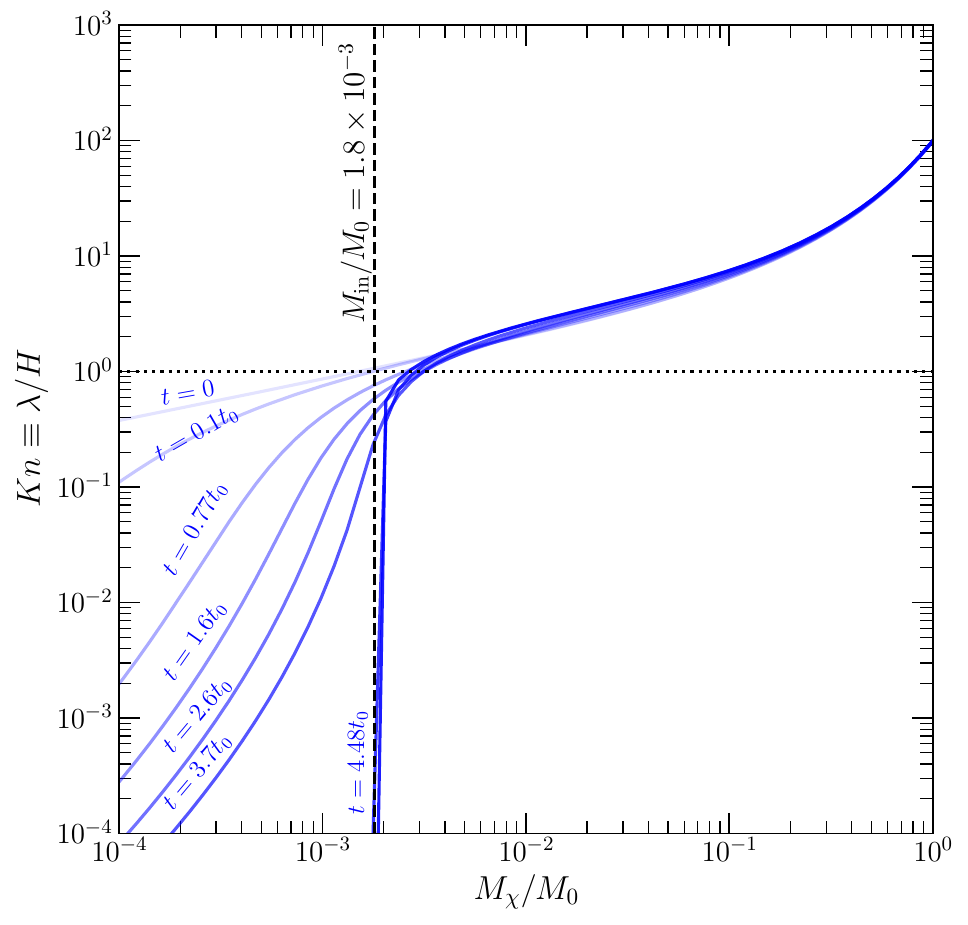}
   \caption{Left: Evolution of dark matter mass profiles (solid) with $(\sigma/m)r_s\rho_s=0.5$, together with the fixed baryon mass profile (dash-dotted). Each dark matter profile is labeled with its corresponding evolution time. The dashed line indicates the mass of the central halo with $Kn<1$. Right: Corresponding $Kn$ value vs. enclosed mass. The dotted horizontal line indicates $Kn=1$, the boundary between short- and long-mean-free-path regimes, where $Kn<1$ and $>1$, respectively.}
   \label{fig:mass}
\end{figure*}

The procedure of our semi-analytical simulations is largely based on the treatment given in~\cite{Balberg:2002ue,Pollack:2014rja, Essig:2018pzq}. We first translate a relevant physical quantity $x$ to a dimensionless one $\hat x$ as $\hat x = x/x_0$, where $x_0$ is its corresponding fiducial value built from the halo parameters $\rho_s$ and $r_s$, as shown in Table~\ref{tab:fiducial}.
\begin{table}[h]
   \centering
   \begin{tabular}{@{} |cc| @{}} 
      \hline
       $M_0 = 4\pi  \rho_s r_s^3$ &  $(\sigma/m)_0 = ({r_s \rho_s})^{-1}$     \\
       $\nu_0 = (4\pi G \rho_s)^{1/2}  r_s$  & $L_0 = (4\pi)^{5/2} G^{3/2} \rho_s^{5/2} r_s^5\label{eq:L0}$  \\
$t_{0} =(4\pi G \rho_s)^{-1/2}$ &  \\ 
      \hline
   \end{tabular}
         \caption{Fiducial quantities used in our numerical simulations.}
   \label{tab:fiducial}
\end{table}

The self-gravitating halo is segmented to $N=182$ evenly log-spaced concentric shells in radius $\{\hat r_1, \hat r_2,\cdots,\hat r_N\}$ with $\hat r_1 = 10^{-4}$ and $\hat r_N = 100$. The halo is assumed to be in a quasi-hydrostatic equilibrium and each shell is assumed to be in its local thermal equilibrium. The values of extensive quantities ($\hat M_i$, $\hat L_i$) and intensive quantities ($\hat \rho_i$, $\hat \nu_i$) are taken as the value at $\hat r_i$ and the average between values at $\hat r_i$ and $\hat r_{i-1}$, respectively. We fix the baryon mass profile $\hat M_{b,i}$ as 
\beq
\hat M_{b,i} =\hat M_b (\hat r_i) = 0.1 \times \hat r_i^{0.6}.
\label{eq:dimlesssingle}
\eeq
Consequently, we only use \emph{one} set of Lagrangian zone radius for the halo through the simulations and dynamically update the enclosed baryon mass according to~\eqref{eq:dimlesssingle}. The workflow is as follows: 

\begin{enumerate}
\item Compute the initial 1D velocity dispersion profile $\hat \nu_{\chi,i}$ based on the input $\hat r_i$, $\hat \rho_{\chi,i}$, and $\hat \rho_{b,i}$ under the hydrostatic equilibrium condition, 
\beq
\f{\partial (\hat \rho_\chi \hat \nu_\chi^2)}{\partial \hat r} = - \f{(\hat M_\chi+\hat M_b) \hat \rho_\chi}{\hat r^2}.
\eeq 

\item Compute the luminosity profile $\hat L_{\chi,i}$ based on $\hat r_i$, $\hat \rho_{\chi,i}$, $\hat \nu_{\chi,i}$ and $\hat \sigma$ according to Eq.~(2) of the main text.

\item Allow a small passage of time $\Delta \hat t$ and compute the specific energy change $\Delta \hat u_{\chi, i}$, $\hat u_{\chi,i} \equiv 3\hat \nu_{\chi,i}^2/2$, due to heat conduction,
\beq
\f{\Delta \hat u_{\chi,i}}{\Delta \hat t} =  - \left(\frac{\partial \hat L_{\chi}}{\partial \hat M_\chi}\right)_i,
\label{eq:5}
\eeq
where the dark matter density is fixed. We then update $\hat u_{\chi,i}$ with $\hat u_{\chi,i}+\Delta \hat u_{\chi, i}$. The time step $\Delta \hat t$ is sufficiently small, i.e., $|\Delta \hat u_{\chi, i}/\hat u_{\chi, i}| < 10^{-4}$.
such that the linear approximations used in step 4 below are valid.

\item Upon updating $\hat u_{\chi, i}$, the $i$-th dark matter halo shell is no longer virialized. To return to hydrostatic equilibrium, we perturb $\hat r_i$, $\hat \rho_{\chi,i}$, and $\hat \nu_{\chi,i}$, while keeping the mass $\hat M_{\chi, i}$ and specific entropy $\hat s_{\chi,i} = \ln (\hat \nu_{\chi,i}^3/\hat \rho_{\chi,i})$ of the shell fixed. We treat mass conservation, specific entropy conservation, kinetic energy conservation, and hydrostatic equilibrium relations, shown in the main text, at the linear order and solve them for all shells simultaneously. For the hydrostatic equilibrium relation, we take the sum of $\hat M_{\chi, i}$ and $\hat M_{b, i}=\hat M_b (\hat r_i)$ to compute the gravitational potential. For numerical accuracy, we iteratively perform the perturbation $10$ times until hydrostatic equilibrium is established everywhere.  

\item Re-establishing hydrostatic equilibrium gives new values for $\hat r_i$, $\hat \rho_{\chi, i}$, and $\hat \nu_{\chi, i}$. We return to step 1 and update the luminosity $\hat L_i$.

\item Track the Knudsen number $Kn\equiv \lambda/H$ for the innermost shell. The evolution is terminated when $Kn$ drops below $10^{-4}$.

\end{enumerate}
The above procedure is coded in \texttt{C++} with the \texttt{eigen 3.2} library for linear algebra~\citep{eigenweb}. 

In~\figref{mass}, we show evolution of  dark matter mass profile (left panel) and the corresponding $Kn$ value vs. enclosed mass (right panel). These results are complementary to those presented in Fig.~1 of the main text.

\section{Angular momentum dissipation}

Dark matter self-interactions provide an important avenue to transport angular momentum. To estimate this effect, we keep track of the collapsing central halo in a Lagrangian zone manner, i.e., the number of particles in each mass sphere is conserved, while allowing its corresponding radius to change over the evolution. This is consistent with the method we use for solving the conduction fluid equations. We further assume the mass distribution is spherically symmetric, which is particularly motivated in SIDM~\citep{Dave:2000ar,Peter:2012jh}, and write the moment of inertia as $I_\chi=k M_{\rm in}r^2_{\rm in}$, where $k=2(\beta+3)/3(\beta+5)$ for a power law of $\rho_\chi\propto r^{\beta}$. For an NFW profile, $\beta= -1$ and $k = 1/3$. For a density core, we have $\beta=0$ and $k=2/5$. The angular momentum is given by
\beq
{J_{\chi, \rm in}}=I_\chi\omega=k{M_{\rm in}}r_{\rm in}^2\omega\simeq \text{const.},
\eeq
where $\omega$ is the rotational frequency of the inner region, $r_{\rm in}$ is its boundary that changes with time. This leads to
\beq
\frac{d}{dt}\left(r_{\rm in}\omega\right)=\frac{d}{dt}\left(\frac{{J_{\chi, \rm in}}}{k{M_{\rm in}}r_{\rm in}}\right)\simeq-\frac{{J_{\chi, \rm in}}}{k{M_{\rm in}}r_{\rm in}^2}\left(\frac{dr_{\rm in}}{dt}\right).
\label{eq:CR}
\eeq
The bulk velocity is $v_{\phi}=r_{\rm in}\omega\sin\theta$ along $\phi$-direction of the rotational axis. $\theta$ is the polar angle. The bulk velocity increases (decreases) $dv_{\phi}/dt>0~(<0)$ as the boundary of the inner region shrinks (expands) $d r_{\rm in}/dt<0~(>0)$. An increasing in $v_\phi$ will drag the ambient regions just outside the boundary and exert a shear pressure on the boundary bulk surface,
\beq
\frac{1}{A_{\rm in}}\frac{d}{dt}\left(N_{r_\text{in}} m v_\phi\right)_{\pm}=\mp\eta_{r_\text{in}}\frac{dv_\phi}{d r_\text{in}}
\label{eq:CA}
\eeq
where $m$ is the dark matter particle mass, $A_\text{in}=4\pi r_\text{in}^2$ is the surface area of the inner region, $\eta_{r_\text{in}}\equiv\eta(r_\text{in})$ is the viscosity of the SIDM fluid, and $N_{r_\text{in}}$ is number of particles, on the bulk surface. The subscript $+/-$ indicates if the quantity increases/decreases. Given the Lagrangian zone setup, $N_{r_\text{in}}$ is a constant through the evolution, and from \eqref{eq:CA} we can show
\beq
\left(\frac{d{r_{\rm in}}}{dt}\right)_{\mp}=\mp\frac{4\pi {r^2_{\rm in}}\eta_{r_{\rm in}}}{N_{r_{\rm in}} m}
\label{eq:CC}
\eeq
The bulk momentum can be transported out through the shear pressure due to viscosity. Combining~\eqref{eq:CR} and~\eqref{eq:CC}, we obtain the rate of momentum transport to the surroundings
\begin{align*}
\frac{d}{dt}\left( {r_{\rm in}}\omega\right)_{\pm}=\pm\frac{4 \pi \eta_{r_{\rm in}} {J_{\chi, \rm in}}}{{kM_{\rm in}}N_{r_{\rm in}} m}
\end{align*}
As the total angular momentum is conserved, the {\em loss} of the angular momentum of the inner region (shrinking case) is given by
\begin{align}
\frac{d{J_{\chi, \rm in}}}{dt}\simeq{}& -N_{r_{\rm in}} m\int\frac{d\Omega}{4\pi} [{r_{\rm in}}\sin\theta\times\frac{d}{dt}\left({r_{\rm in}}\omega\sin\theta\right)]\nonumber\\
={}&-\int\frac{\sin^3\theta d\theta d\phi}{4\pi}{r_{\rm in}}\frac{d}{dt}\left(N_{r_{\rm in}} m {r_{\rm in}}\omega\right)_{+} \nonumber\\
={}&-\frac{8\pi}{3}\frac{J_{\chi, \rm in}}{{kM_{\rm in}}}\eta_{r_{\rm in}} {r_{\rm in}}.
\end{align}
We have
\begin{equation}
{J_{\chi, \rm in}}(t; M_{\rm in})=J_{\chi, \rm in}^{i}\exp\left(-\frac{8\pi}{3}\int_{t_i}^{t}\frac{\eta_{r_{\rm in}} r_{\rm in}(t')}{kM_{\rm in}}dt'\right),
\end{equation}
where ${J}_{\chi,\rm in}^{i}$ is the initial angular momentum of the inner region.
As for conductivity~\cite{Balberg:2002ue}, the viscosity for both long-mean-free-path and short-mean-free-path regimes can be combined into a single expression,
\begin{align}
\eta={}&\frac{1}{3}mn\bar{v}\left(\frac{1}{\lambda}+\frac{\nu t_r}{H^2}\right)^{-1} \nonumber\\
={}&\frac{1}{3}\alpha(\sigma/m)\bar{v}\left[\alpha(\sigma/m)^2+\frac{4\pi G}{\rho \nu^2}\right]^{-1},
\end{align}
where we have used the gravitational scale height $H=\sqrt{\nu^2/4\pi G\rho}$, the mean free path $\lambda=1/n\sigma$ and the relaxation time $t_r=1/(\alpha n \nu\sigma)$ with number density $n$, cross section $\sigma$, and $\alpha=(16/\pi)^{1/2}\approx 2.26$ for hard spheres.

We evaluate $\eta$ at the boundary $r_{\rm in}$ and take $\bar{v}\simeq \sqrt{3} \nu$, and obtain 
\beq
\label{eq:long}
{J_{\chi, \rm in}}=J_{\chi, \rm in}^{i}\exp\left[-\frac{8}{\sqrt{27\pi}}\int_{t_i}^{t}\frac{\rho_{r_{\rm in}}(t')(\sigma/m) \nu_{r_{\rm in}}^{3}(t')r_{\rm in}(t')}{kG{M_{\rm in}}}dt'\right]
\eeq
in the long-mean-free-path limit, and 
\beq
{J_{\chi, \rm in}}=J_{\chi, \rm in}^{i}\exp\left[-\frac{8\pi}{3\sqrt{3}}\int_{t_i}^{t}\frac{\nu_{r_{\rm in}}(t') r_{\rm in}(t')}{{kM_{\rm in}}(\sigma/m)}dt'\right]
\eeq
in the short-mean-free-path limit, where we have used the density $\rho_{r_{\rm in}}=\rho(r_{\rm in})=mn(r_{\rm in})$ and $\nu_{r_{\rm in}}=\nu(r_{\rm in})$ the 1D velocity dispersion at the boundary. 

For the benchmark discussed in Section 6, we estimate the characteristic timescale to dissipate the angular momentum remnant $\Delta t\sim0.1t_0$ with fixed $M_{\rm in}=1.8\times10^{-3}M_0$ and its corresponding $r_{\rm in}=0.063r_s$, i.e., the radius at which the enclosed mass of the initial NFW halo is $M_{\rm in}$. We do not take into account the change in the radius of the mass sphere and the velocity dispersion, as they are negligible for $\Delta t\sim0.1t_0$; see the left panel of Fig.~\ref{fig:mass}. The estimation is based on Eq.~(\ref{eq:long}), and this is well justified as the collapsed region is in the long-mean-free-path regime at $r\sim r_{\rm in}$ for $\Delta t\sim0.1t_0$, as shown in the right panel of Fig.~\ref{fig:mass}. For a sphere with smaller $M_{\rm in}$ more towards the center, the timescale is even shorter, as we can see from the scaling relation $\Delta t\sim GM_\chi/(\rho_\chi\nu^3_\chi r\sigma/m)$. For an NFW halo, we have $\rho_\chi\propto r^{-1}$, $M_\chi\propto r^{2}$, $\nu_\chi\propto r^{0.2}$, and hence $\Delta t\propto r^{1.4}$. For a cored halo, $\rho_\chi\propto r^{0}$, $M_\chi\propto r^{3}$, $\nu_\chi\propto r^{0}$, and $\Delta t\propto r^{2}$. For both cases, $\Delta t$ decreases as $r$ reduces. Thus it's reasonable to collectively treat particles in the mass sphere $M_{\rm in}=1.8\times10^{-3}M_0$, which would collapse to a seed, and estimate the overall timescale for dissipating angular momentum.

\begin{table*}[htbp]
   \centering
   \begin{tabular}{@{} ccrrrc @{}}
         \hline
         Label & Name & $M_\text{BH}$ [$10^9 \text{M}_\odot$]   & $z$ & $f_\text{Edd}$ & Ref.\\
      \hline
         1  & J1205$-0000$ & $2.2^{+0.2}_{-0.6}$ & $6.699^{+0.007}_{-0.001}$ & $0.16^{+0.04}_{-0.02}$&\cite{onoue2019subaru}\\
      2 &J1243$+0100$ & $0.33^{+0.2}_{-0.2}$ & $7.07^{+0.01}_{-0.01}$ & $0.34^{+0.02}_{-0.02}$ &\cite{matsuoka2019discovery}\\
      3 & J2239$+0207$ & $1.1^{+0.3}_{-0.2}$ & $6.245^{+0.008}_{-0.007}$ & $0.17^{+0.04}_{-0.05}$&\cite{onoue2019subaru} \\
      4 & J2216$-0016$ & $0.7^{+0.14}_{-0.23}$ & $6.109^{+0.007}_{-0.008}$ &  $0.15^{+0.05}_{-0.03}$ & \cite{onoue2019subaru} \\
      5 & J1208$-0200$ & $0.71^{+0.24}_{-0.52}$ & $6.144^{+0.008}_{-0.010}$ & $0.24^{+0.18}_{-0.08}$ & \cite{onoue2019subaru}\\
        6     & J1007$+2115$ & $1.5^{+0.2}_{-0.2} $ & $7.5149^{+0.0004}_{-0.0004}$ & $1.06^{+0.2}_{-0.2}$ &\cite{yang2020poniua}\\
      7     & J1342$+0928$ & $0.78^{+0.33}_ {- 0.19}$ & $7.5413^{+0.0007}_{-0.0007}$ & $1.5^{+0.5}_{-0.4}$ & \cite{Banados:2017unc}\\
      8      & J1120$+0641$ & $2.0^{+1.5}_{-0.7}$ & $7.085^{+0.003}_{-0.003}$ & $1.2^{+0.6}_{-0.5}$ &\cite{mortlock2011luminous}\\
      9  & J0038$-1527$ & $1.33^{+0.25}_{-0.25}$ & $7.021^{+0.005}_{-0.005}$ & $1.25^{+0.19}_{-0.19}$ &\cite{wang2018discovery}\\
      10 & J0100$+2802$ & $12.4^{+1.9}_{-1.9}$ & $6.30^{+0.01}_{-0.01}$ & $0.99^{+0.22}_{-0.22}$ & \cite{wu2015ultraluminous}\\
      \hline
   \end{tabular}
         \caption{The sample of high-$z$ SMBHs shown in Fig. 2 of the main text.}
   \label{tab:example}
\end{table*}

\section{The adiabatic index}

We consider a perfect fluid with energy density $\rho(r)c^2$ and pressure $p(r)$ in a Schwarzschild metric~\citep{Chandrasekhar:1964zz}
\[
ds^2=-e^{2\Phi(r)}c^2dt^2+e^{2\Lambda(r)}dr^2+r^2(d\theta^2+\sin^2\theta~d\phi^2),
\]
where $e^{2\Phi}=\exp[2\int_r^\infty (dp/dr')/(p+\rho c^2)dr']$ and $e^{2\Lambda}=[1-2GM(r)/rc^2]^{-1}$.
The critical adiabatic index is 
\begin{align}
\gamma_{\text{cr}}\equiv&\frac{4}{3}+\frac{\int e^{3\Phi+\Lambda}[16p+(e^{2\Lambda}-1)(\rho+p)](e^{2\Lambda}-1)r^2dr}{36\int e^{3\Phi+\Lambda}pr^2dr}\nonumber\\
&+\frac{4\pi G\int e^{3(\Phi+\Lambda)}[8p+(e^{2\Lambda}+1)(\rho c^2+p)]pr^4dr}{9c^4\int e^{3\Phi+\Lambda}pr^2dr}\nonumber\\
&+\frac{16\pi^2 G^2\int e^{3\Phi+5\Lambda}(\rho c^2+p)p^2r^6dr}{9c^8\int e^{3\Phi+\Lambda}pr^2dr}
\end{align}
and the pressure-averaged adiabatic index is $\langle\gamma\rangle\equiv{\int e^{3\Phi+\Lambda}\gamma(r) pr^2dr}/({\int e^{3\Phi+\Lambda}pr^2dr})$. These expressions are fully relativistic, and we will provide their derivations in a companion paper~\citep{fyz}.

\section{The sample of high-$z$ SMBHs}

In Table~\ref{tab:example}, we list high-$z$ SMBHs shown in Fig. 2 of the main text, in the order of their labeling number in the figure. The Eddington ratio is calculated as $f_\text{Edd}=L_\text{bol}/L_\text{Edd}$, where $L_\text{bol}$ is the observed bolometric luminosity and $L_\text{Edd} = 1.3\times 10^{38} (M_\text{BH}/{{\rm M}_\odot})~\text{erg/s}$ is the Eddington luminosity.

\bibliography{smbh}{}

\begin{thebibliography}{}
\expandafter\ifx\csname natexlab\endcsname\relax\def\natexlab#1{#1}\fi
\providecommand{\url}[1]{\href{#1}{#1}}
\providecommand{\dodoi}[1]{doi:~\href{http://doi.org/#1}{\nolinkurl{#1}}}
\providecommand{\doeprint}[1]{\href{http://ascl.net/#1}{\nolinkurl{http://ascl.net/#1}}}
\providecommand{\doarXiv}[1]{\href{https://arxiv.org/abs/#1}{\nolinkurl{https://arxiv.org/abs/#1}}}

\bibitem[{Aghanim {et~al.}(2020)}]{Aghanim:2018eyx}
Aghanim, N., {et~al.} 2020, Astron. Astrophys., 641, A6,
  \dodoi{10.1051/0004-6361/201833910}

\bibitem[{Ahn \& Shapiro(2005)}]{Ahn:2004xt}
Ahn, K.-J., \& Shapiro, P.~R. 2005, Mon. Not. Roy. Astron. Soc., 363, 1092,
  \dodoi{10.1111/j.1365-2966.2005.09492.x}

\bibitem[{Alexander \& Natarajan(2014)}]{Alexander:2014noa}
Alexander, T., \& Natarajan, P. 2014, Science, 345, 1330,
  \dodoi{10.1126/science.1251053}

\bibitem[{Andrade {et~al.}(2021)Andrade, Fuson, Gad-Nasr, Kong, Minor, Roberts,
  \& Kaplinghat}]{Andrade:2020lqq}
Andrade, K.~E., Fuson, J., Gad-Nasr, S., {et~al.} 2021, Mon. Not. Roy. Astron.
  Soc., 510, 54, \dodoi{10.1093/mnras/stab3241}

\bibitem[{Balberg \& Shapiro(2002)}]{Balberg:2001qg}
Balberg, S., \& Shapiro, S.~L. 2002, Phys.Rev.Lett., 88, 101301,
  \dodoi{10.1103/PhysRevLett.88.101301}

\bibitem[{Balberg {et~al.}(2002)Balberg, Shapiro, \& Inagaki}]{Balberg:2002ue}
Balberg, S., Shapiro, S.~L., \& Inagaki, S. 2002, Astrophys. J., 568, 475,
  \dodoi{10.1086/339038}

\bibitem[{Banados {et~al.}(2018)}]{Banados:2017unc}
Banados, E., {et~al.} 2018, Nature, 553, 473, \dodoi{10.1038/nature25180}

\bibitem[{{Begelman}(1979)}]{Begelman:1979}
{Begelman}, M.~C. 1979, \mnras, 187, 237, \dodoi{10.1093/mnras/187.2.237}

\bibitem[{Begelman {et~al.}(2006)Begelman, Volonteri, \&
  Rees}]{Begelman:2006db}
Begelman, M.~C., Volonteri, M., \& Rees, M.~J. 2006, Mon. Not. Roy. Astron.
  Soc., 370, 289, \dodoi{10.1111/j.1365-2966.2006.10467.x}

\bibitem[{Bromm \& Loeb(2003)}]{Bromm:2002hb}
Bromm, V., \& Loeb, A. 2003, Astrophys. J., 596, 34, \dodoi{10.1086/377529}

\bibitem[{Chandrasekhar(1964)}]{Chandrasekhar:1964zz}
Chandrasekhar, S. 1964, Astrophys. J., 140, 417, \dodoi{10.1086/147938}

\bibitem[{Choquette {et~al.}(2019)Choquette, Cline, \&
  Cornell}]{Choquette:2018lvq}
Choquette, J., Cline, J.~M., \& Cornell, J.~M. 2019, JCAP, 1907, 036,
  \dodoi{10.1088/1475-7516/2019/07/036}

\bibitem[{Cruz {et~al.}(2020)Cruz, Pontzen, Volonteri, Quinn, Tremmel, Brooks,
  Sanchez, Munshi, \& Cintio}]{Cruz:2020rit}
Cruz, A., Pontzen, A., Volonteri, M., {et~al.} 2020, Mon. Not. Roy. Astron.
  Soc., 500, 2177, \dodoi{10.1093/mnras/staa3389}

\bibitem[{Dave {et~al.}(2001)Dave, Spergel, Steinhardt, \&
  Wandelt}]{Dave:2000ar}
Dave, R., Spergel, D.~N., Steinhardt, P.~J., \& Wandelt, B.~D. 2001, Astrophys.
  J., 547, 574, \dodoi{10.1086/318417}

\bibitem[{Dutton \& Macci{\`o}(2014)}]{Dutton:2014xda}
Dutton, A.~A., \& Macci{\`o}, A.~V. 2014, Mon. Not. Roy. Astron. Soc., 441,
  3359, \dodoi{10.1093/mnras/stu742}

\bibitem[{Essig {et~al.}(2019)Essig, Mcdermott, Yu, \& Zhong}]{Essig:2018pzq}
Essig, R., Mcdermott, S.~D., Yu, H.-B., \& Zhong, Y.-M. 2019, Phys. Rev. Lett.,
  123, 121102, \dodoi{10.1103/PhysRevLett.123.121102}

\bibitem[{Feng {et~al.}(2021)Feng, Yu, \& Zhong}]{fyz}
Feng, W.-X., Yu, H.-B., \& Zhong, Y.-M. 2021, {In Preparation}

\bibitem[{Freese {et~al.}(2010)Freese, Ilie, Spolyar, Valluri, \&
  Bodenheimer}]{Freese:2010re}
Freese, K., Ilie, C., Spolyar, D., Valluri, M., \& Bodenheimer, P. 2010,
  Astrophys. J., 716, 1397, \dodoi{10.1088/0004-637X/716/2/1397}

\bibitem[{Guennebaud {et~al.}(2010)Guennebaud, Jacob, {et~al.}}]{eigenweb}
Guennebaud, G., Jacob, B., {et~al.} 2010, Eigen v3, http://eigen.tuxfamily.org

\bibitem[{Hernquist(1990)}]{Hernquist:1990be}
Hernquist, L. 1990, Astrophys. J., 356, 359, \dodoi{10.1086/168845}

\bibitem[{Huo {et~al.}(2020)Huo, Yu, \& Zhong}]{Huo:2019yhk}
Huo, R., Yu, H.-B., \& Zhong, Y.-M. 2020, JCAP, 2006, 051,
  \dodoi{10.1088/1475-7516/2020/06/051}

\bibitem[{Inayoshi {et~al.}(2020)Inayoshi, Visbal, \&
  Haiman}]{Inayoshi:2019fun}
Inayoshi, K., Visbal, E., \& Haiman, Z. 2020, Ann. Rev. Astron. Astrophys., 58,
  27, \dodoi{10.1146/annurev-astro-120419-014455}

\bibitem[{Kaplinghat {et~al.}(2020)Kaplinghat, Ren, \& Yu}]{Kaplinghat:2019dhn}
Kaplinghat, M., Ren, T., \& Yu, H.-B. 2020, JCAP, 06, 027,
  \dodoi{10.1088/1475-7516/2020/06/027}

\bibitem[{Kaplinghat {et~al.}(2016)Kaplinghat, Tulin, \&
  Yu}]{Kaplinghat:2015aga}
Kaplinghat, M., Tulin, S., \& Yu, H.-B. 2016, Phys. Rev. Lett., 116, 041302,
  \dodoi{10.1103/PhysRevLett.116.041302}

\bibitem[{Kerr(1963)}]{Kerr:1963ud}
Kerr, R.~P. 1963, Phys. Rev. Lett., 11, 237, \dodoi{10.1103/PhysRevLett.11.237}

\bibitem[{King(1966)}]{King:1966fn}
King, I.~R. 1966, Astron. J., 71, 64, \dodoi{10.1086/109857}

\bibitem[{Koda \& Shapiro(2011)}]{Koda:2011yb}
Koda, J., \& Shapiro, P.~R. 2011, Mon. Not. Roy. Astron. Soc., 415, 1125,
  \dodoi{10.1111/j.1365-2966.2011.18684.x}

\bibitem[{Kulkarni {et~al.}(2019)Kulkarni, Worseck, \&
  Hennawi}]{Kulkarni:2018ebj}
Kulkarni, G., Worseck, G., \& Hennawi, J.~F. 2019, Mon. Not. Roy. Astron. Soc.,
  488, 1035, \dodoi{10.1093/mnras/stz1493}

\bibitem[{Liao {et~al.}(2017)Liao, Chen, \& Chu}]{Liao:2016tdi}
Liao, S., Chen, J., \& Chu, M.-C. 2017, Astrophys. J., 844, 86,
  \dodoi{10.3847/1538-4357/aa79fb}

\bibitem[{Lynden-Bell \& Wood(1968)}]{LyndenBell:1968yw}
Lynden-Bell, D., \& Wood, R. 1968, Mon. Not. Roy. Astron. Soc., 138, 495

\bibitem[{Matsuoka {et~al.}(2019)Matsuoka, Onoue, Kashikawa, Strauss, Iwasawa,
  Lee, Imanishi, Nagao, Akiyama, Asami, {et~al.}}]{matsuoka2019discovery}
Matsuoka, Y., Onoue, M., Kashikawa, N., {et~al.} 2019, Astrophys. J. Lett.,
  872, L2

\bibitem[{McKinney {et~al.}(2014)McKinney, Tchekhovskoy, Sadowski, \&
  Narayan}]{McKinney:2013txa}
McKinney, J.~C., Tchekhovskoy, A., Sadowski, A., \& Narayan, R. 2014, Mon. Not.
  Roy. Astron. Soc., 441, 3177, \dodoi{10.1093/mnras/stu762}

\bibitem[{{Merafina} \& {Ruffini}(1989)}]{Merafina:1989}
{Merafina}, M., \& {Ruffini}, R. 1989, Astron. Astrophys., 221, 4

\bibitem[{Mo {et~al.}(2010)Mo, van~den Bosch, \& White}]{mo2010galaxy}
Mo, H., van~den Bosch, F., \& White, S. 2010, Galaxy Formation and Evolution,
  Galaxy Formation and Evolution (Cambridge University Press)

\bibitem[{Mortlock {et~al.}(2011)Mortlock, Warren, Venemans, Patel, Hewett,
  McMahon, Simpson, Theuns, Gonz{\'a}les-Solares, Adamson,
  {et~al.}}]{mortlock2011luminous}
Mortlock, D.~J., Warren, S.~J., Venemans, B.~P., {et~al.} 2011, Nature, 474,
  616

\bibitem[{Natarajan(2011)}]{Natarajan2011}
Natarajan, P. 2011, Bull. Astr. Soc. India, 39, 145

\bibitem[{Navarro {et~al.}(1996)Navarro, Frenk, \& White}]{Navarro:1995iw}
Navarro, J.~F., Frenk, C.~S., \& White, S.~D. 1996, Astrophys. J., 462, 563,
  \dodoi{10.1086/177173}

\bibitem[{Onoue {et~al.}(2019)Onoue, Kashikawa, Matsuoka, Kato, Izumi, Nagao,
  Strauss, Harikane, Imanishi, Ito, {et~al.}}]{onoue2019subaru}
Onoue, M., Kashikawa, N., Matsuoka, Y., {et~al.} 2019, Astrophys. J., 880, 77

\bibitem[{Peter {et~al.}(2013)Peter, Rocha, Bullock, \&
  Kaplinghat}]{Peter:2012jh}
Peter, A. H.~G., Rocha, M., Bullock, J.~S., \& Kaplinghat, M. 2013, Mon. Not.
  Roy. Astron. Soc., 430, 105, \dodoi{10.1093/mnras/sts535}

\bibitem[{Pollack {et~al.}(2015)Pollack, Spergel, \&
  Steinhardt}]{Pollack:2014rja}
Pollack, J., Spergel, D.~N., \& Steinhardt, P.~J. 2015, Astrophys. J., 804,
  131, \dodoi{10.1088/0004-637X/804/2/131}

\bibitem[{Press \& Schechter(1974)}]{Press:1973iz}
Press, W.~H., \& Schechter, P. 1974, Astrophys. J., 187, 425,
  \dodoi{10.1086/152650}

\bibitem[{Ren {et~al.}(2019)Ren, Kwa, Kaplinghat, \& Yu}]{Ren:2018jpt}
Ren, T., Kwa, A., Kaplinghat, M., \& Yu, H.-B. 2019, Phys. Rev. X, 9, 031020,
  \dodoi{10.1103/PhysRevX.9.031020}

\bibitem[{Robertson {et~al.}(2018)}]{Robertson:2017mgj}
Robertson, A., {et~al.} 2018, Mon. Not. Roy. Astron. Soc., 476, L20,
  \dodoi{10.1093/mnrasl/sly024}

\bibitem[{Rocha {et~al.}(2013)Rocha, Peter, Bullock, Kaplinghat,
  Garrison-Kimmel, {et~al.}}]{Rocha:2012jg}
Rocha, M., Peter, A.~H., Bullock, J.~S., {et~al.} 2013,
  Mon.Not.Roy.Astron.Soc., 430, 81, \dodoi{10.1093/mnras/sts514}

\bibitem[{{Sadowski} {et~al.}(2015){Sadowski}, {Narayan}, {Tchekhovskoy},
  {Abarca}, {Zhu}, \& {McKinney}}]{Sadowski:2015}
{Sadowski}, A., {Narayan}, R., {Tchekhovskoy}, A., {et~al.} 2015, \mnras, 447,
  49, \dodoi{10.1093/mnras/stu2387}

\bibitem[{Saijo {et~al.}(2002)Saijo, Baumgarte, Shapiro, \&
  Shibata}]{Saijo:2002qt}
Saijo, M., Baumgarte, T.~W., Shapiro, S.~L., \& Shibata, M. 2002, Astrophys.
  J., 569, 349, \dodoi{10.1086/339268}

\bibitem[{Salpeter(1964)}]{Salpeter:1964kb}
Salpeter, E. 1964, Astrophys. J., 140, 796, \dodoi{10.1086/147973}

\bibitem[{Sameie {et~al.}(2018)Sameie, Creasey, Yu, Sales, Vogelsberger, \&
  Zavala}]{Sameie:2018chj}
Sameie, O., Creasey, P., Yu, H.-B., {et~al.} 2018, Mon. Not. Roy. Astron. Soc.,
  479, 359, \dodoi{10.1093/mnras/sty1516}

\bibitem[{Sameie {et~al.}(2020)Sameie, Yu, Sales, Vogelsberger, \&
  Zavala}]{Sameie:2019zfo}
Sameie, O., Yu, H.-B., Sales, L.~V., Vogelsberger, M., \& Zavala, J. 2020,
  Phys. Rev. Lett., 124, 141102, \dodoi{10.1103/PhysRevLett.124.141102}

\bibitem[{Sameie {et~al.}(2021)Sameie, Boylan-Kolchin, Sanderson, Vargya,
  Hopkins, Wetzel, Bullock, Graus, \& Robles}]{Sameie:2021ang}
Sameie, O., Boylan-Kolchin, M., Sanderson, R., {et~al.} 2021, Mon. Not. Roy.
  Astron. Soc., 507, 720, \dodoi{10.1093/mnras/stab2173}

\bibitem[{Shakura \& Sunyaev(1976)}]{Shakura:1976xk}
Shakura, N.~I., \& Sunyaev, R. 1976, Mon. Not. Roy. Astron. Soc., 175, 613

\bibitem[{Shapiro \& Teukolsky(1983)}]{Shapiro:1983du}
Shapiro, S., \& Teukolsky, S. 1983, {Black holes, white dwarfs, and neutron
  stars: The physics of compact objects} (New York: John Wiley and Sons)

\bibitem[{Shapiro(2005)}]{Shapiro:2004ud}
Shapiro, S.~L. 2005, Astrophys. J., 620, 59, \dodoi{10.1086/427065}

\bibitem[{Spergel \& Steinhardt(2000)}]{Spergel:1999mh}
Spergel, D.~N., \& Steinhardt, P.~J. 2000, Phys. Rev. Lett., 84, 3760,
  \dodoi{10.1103/PhysRevLett.84.3760}

\bibitem[{Trakhtenbrot(2020)}]{Trakhtenbrot:2020ugp}
Trakhtenbrot, B. 2020, in {IAU Symposium 356}: {Nuclear Activity in Galaxies
  Across Cosmic Time}.
\newblock \doarXiv{2002.00972}

\bibitem[{Tulin \& Yu(2018)}]{Tulin:2017ara}
Tulin, S., \& Yu, H.-B. 2018, Phys. Rept., 730, 1,
  \dodoi{10.1016/j.physrep.2017.11.004}

\bibitem[{Vogelsberger {et~al.}(2012)Vogelsberger, Zavala, \&
  Loeb}]{Vogelsberger:2012ku}
Vogelsberger, M., Zavala, J., \& Loeb, A. 2012, Mon. Not. Roy. Astron. Soc.,
  423, 3740, \dodoi{10.1111/j.1365-2966.2012.21182.x}

\bibitem[{Vogelsberger {et~al.}(2014)Vogelsberger, Zavala, Simpson, \&
  Jenkins}]{Vogelsberger:2014vr}
Vogelsberger, M., Zavala, J., Simpson, C., \& Jenkins, A. 2014, Mon. Not. Roy.
  Astron. Soc., 444, 3684, \dodoi{10.1093/mnras/stu1713}

\bibitem[{Volonteri(2010)}]{Volonteri:2010wz}
Volonteri, M. 2010, Astron. Astrophys. Rev., 18, 279,
  \dodoi{10.1007/s00159-010-0029-x}

\bibitem[{Volonteri \& Rees(2005)}]{Volonteri:2005fj}
Volonteri, M., \& Rees, M.~J. 2005, Astrophys. J., 633, 624,
  \dodoi{10.1086/466521}

\bibitem[{Wang {et~al.}(2018)Wang, Yang, Fan, Yue, Wu, Schindler, Bian, Li,
  Farina, Ba{\~n}ados, {et~al.}}]{wang2018discovery}
Wang, F., Yang, J., Fan, X., {et~al.} 2018, Astrophys. J. Lett., 869, L9

\bibitem[{Wise {et~al.}(2008)Wise, Turk, \& Abel}]{Wise:2007bf}
Wise, J.~H., Turk, M.~J., \& Abel, T. 2008, Astrophys. J., 682, 745,
  \dodoi{10.1086/588209}

\bibitem[{Wu {et~al.}(2015)Wu, Wang, Fan, Yi, Zuo, Bian, Jiang, McGreer, Wang,
  Yang, {et~al.}}]{wu2015ultraluminous}
Wu, X.-B., Wang, F., Fan, X., {et~al.} 2015, Nature, 518, 512

\bibitem[{Yang \& Yu(2021)}]{Yang:2021kdf}
Yang, D., \& Yu, H.-B. 2021, Phys. Rev. D, 104, 103031,
  \dodoi{10.1103/PhysRevD.104.103031}

\bibitem[{Yang {et~al.}(2020{\natexlab{a}})Yang, Yu, \& An}]{Yang:2020iya}
Yang, D., Yu, H.-B., \& An, H. 2020{\natexlab{a}}, Phys. Rev. Lett., 125,
  111105, \dodoi{10.1103/PhysRevLett.125.111105}

\bibitem[{Yang {et~al.}(2020{\natexlab{b}})Yang, Wang, Fan, Hennawi, Davies,
  Yue, Banados, Wu, Venemans, Barth, {et~al.}}]{yang2020poniua}
Yang, J., Wang, F., Fan, X., {et~al.} 2020{\natexlab{b}}, Astrophys. J. Lett.,
  897, L14

\bibitem[{Zhao {et~al.}(2009)Zhao, Jing, Mo, \& Boerner}]{Zhao:2008wd}
Zhao, D., Jing, Y., Mo, H., \& Boerner, G. 2009, Astrophys. J., 707, 354,
  \dodoi{10.1088/0004-637X/707/1/354}

\end{thebibliography}
\bibliographystyle{aasjournal}

\end{document}